\documentclass{osa-article}

\journal{osajournal}

\articletype{Research Article}

\usepackage{siunitx}
\usepackage{color}

\begin{document}

\title{Disentangling X-ray dichroism and birefringence via high-purity polarimetry}

\author{
Annika T. Schmitt,\authormark{1,2,3,*} 
Yves Joly,\authormark{4}
Kai S. Schulze,\authormark{1,2,3}
Berit Marx-Glowna,\authormark{1,2,3}
Ingo Uschmann, \authormark{1,2,3}
Benjamin Grabiger, \authormark{1,2,3}
Hendrik Bernhardt, \authormark{1,2,3}
Robert Loetzsch, \authormark{1,2,3}
Am\'elie Juhin, \authormark{5}
J\'er\^ome Debray, \authormark{4}
Hans-Christian Wille, \authormark{6}
Hasan Yava\c{s}, \authormark{6,7}
Gerhard G. Paulus, \authormark{1,2,3}
and Ralf R\"ohlsberger \authormark{6,1,2,3,**}
}

\address{
\authormark{1}Institut f\"ur Optik und Quantenelektronik, Friedrich-Schiller-Universit\"at Jena, Max-Wien-Platz 1, 07743 Jena, Germany\\
\authormark{2}Helmholtz-Institut Jena, Fr\"{o}belstieg 3, 07743 Jena, Germany\\
\authormark{3}Helmholtz Centre for Heavy Ion Research (GSI), Planckstr. 1, 64291 Darmstadt, Germany\\
\authormark{4}Institut N\'eel, 25 Rue des Martyrs, BP 166, 38042 Grenoble cedex 09, France\\
\authormark{5}Institut de Min\'eralogie, de Physique des Mat\'eriaux et de Cosmochimie (IMPMC), Sorbonne Universit\'e, UMR 7590, 4 place Jussieu, 75252 Paris Cedex 05, France\\
\authormark{6}Deutsches Elektronen-Synchrotron DESY, Notkestr.\,85, 22607 Hamburg, Germany\\
\authormark{7}SLAC National Accelerator Laboratory, 2575 Sand Hill Road, MS103, Menlo Park, CA 94025

\authormark{*}annika.schmitt@uni-jena.de
\authormark{**}ralf.roehlsberger@desy.de
}




\begin{abstract}
High-brilliance synchrotron radiation sources have opened new avenues for X-ray polarization analysis that go far beyond conventional polarimetry in the optical domain. With linear X-ray polarizers in a crossed setting polarization extinction ratios down to 10$^{-10}$ can be achieved. This renders the method sensitive to probe tiniest optical anisotropies that would occur, for example, in strong-field QED due to vacuum birefringence and dichroism. Here we show that high-purity polarimetry can be employed to reveal electronic anisotropies in condensed matter systems with utmost sensitivity and spectral resolution. Taking CuO and La$_2$CuO$_4$ as benchmark systems, we present a full characterization of the polarization changes across the Cu K-absorption edge and their separation into dichroic and birefringent contributions. At diffraction-limited synchrotron radiation sources and X-ray lasers, where polarization extinction ratios of 10$^{-12}$ can be achieved, our method has the potential to assess birefringence and dichroism of the quantum vacuum in extreme electromagnetic fields.
\end{abstract}

\section{Introduction}
Symmetries in nature are closely related to the fundamental structure of atoms, molecules and solids \cite{Landafshitz1977}.  Symmetry breaking interactions in condensed matter, for example, form the fundamental basis for macroscopic quantum effects like magnetism, superconductivity, giant magnetoresistance, multiferroicity, and others, rendering the optical properties of such materials anisotropic. Access to symmetries and anisotropies of matter has been provided for centuries by optical effects like dichroism and birefringence \cite{BornWolf1999}, in particular by studying how the optical properties of matter depend on the polarization of light and how the polarization of light is affected by the interaction with matter. In recent decades, the use of highly brilliant X-rays from synchrotron radiation sources has provided access to the microscopic origins of magnetic and electronic anisotropies. This is  facilitated, amongst others, via a suite of dichroic X-ray absorption spectroscopies in which the polarization dependence of X-ray absorption in the vicinity of atomic inner-shell transitions is monitored \cite{Templeton_1980, Thole_1985, vanderLaan_1986, Schuetz_1987, Alagna_1998}. 

Polarization changes of X-rays in the interaction with matter occur not only due to absorptive but also through dispersive effects, leading to dichroism and birefringence, respectively. In the case of linearly polarized X-rays, dichroism causes a rotation of the polarization vector due to an anisotropic absorption cross-section of the sample. X-ray birefringence results from different propagation velocities of two orthogonal polarization components, which leads to a phase shift between those and induces an ellipticity of the light. Both effects constitute sensitive probes for fundamental aspects of the light-matter interaction:
In condensed matter physics, the spectral dependencies of X-ray dichroism and birefringence depict a very sensitive fingerprint of the electronic structure of the material. For example, tiny optical anisotropies emerging in the vicinity of phase transitions could reveal precursor mechanisms for structural transformations and electronic ordering in materials \cite{Norman2013,Pershoguba2013,Norman2015}. In quantum electrodynamics (QED) of extremely strong electromagnetic fields it is predicted that even the vacuum becomes optically anisotropic \cite{Heyl1997, Heinzl2006, karbstein2015vacuum, karbstein2018vacuum}. The resulting birefringence and dichroism could be sensitively probed by polarization analysis of hard X-rays after interaction with ultraintense light fields \cite{karbstein2016probing, karbstein2020enhancing, schlenvoigt2016detecting}. This would be a first test of nonlinear QED since the first considerations on this subject by Euler and Heisenberg in 1936 \cite{Heisenberg1936}. It is thus obvious that the precise detection of dichroic and birefringent polarization changes of scattered X-ray radiation would provide fundamental insights into condensed matter physics and QED effects alike. Motivated by these perspectives, very efficient high-purity polarimeters for hard X-rays have been developed with extinction ratios of up to $10^{-10}$ \cite{Marx_2013, Bernhardt2020}. They are based on two crossed linear Bragg polarizers between which the X-ray optical activity takes place \cite{Hart_1978, Siddons_1990, Hart_1991, Siddons_1993, Siddons_1995, Toellner_1995, Roehlsberger_1997, Alp_2000, Heeg_2013, Heeg_2015, Haber_2016}. 

Here we employ high-purity polarimetry to reveal electronic anisotropies in condensed matter with maximum orbital sensitivity, thereby consolidating the research field of spectroscopic polarimetry. We showcase the potential of this analysis technique by presenting a full characterization of the polarization changes across an atomic absorption edge and their separation into dichroic and birefringent contributions.  As benchmark systems for correlated materials and parent compounds for cuprate superconductors, two materials have been chosen, CuO and La$_{\text{2}}$CuO$_{\text{4}}$. We are probing the pronounced electronic anisotropies in these cuprate compounds that result from the particular symmetry of the Cu atom and its hybridizations with the surrounding orbitals in the near-edge region of the Cu K-absorption edge. 

	 \begin{figure}
		\centering\includegraphics[width=12cm]{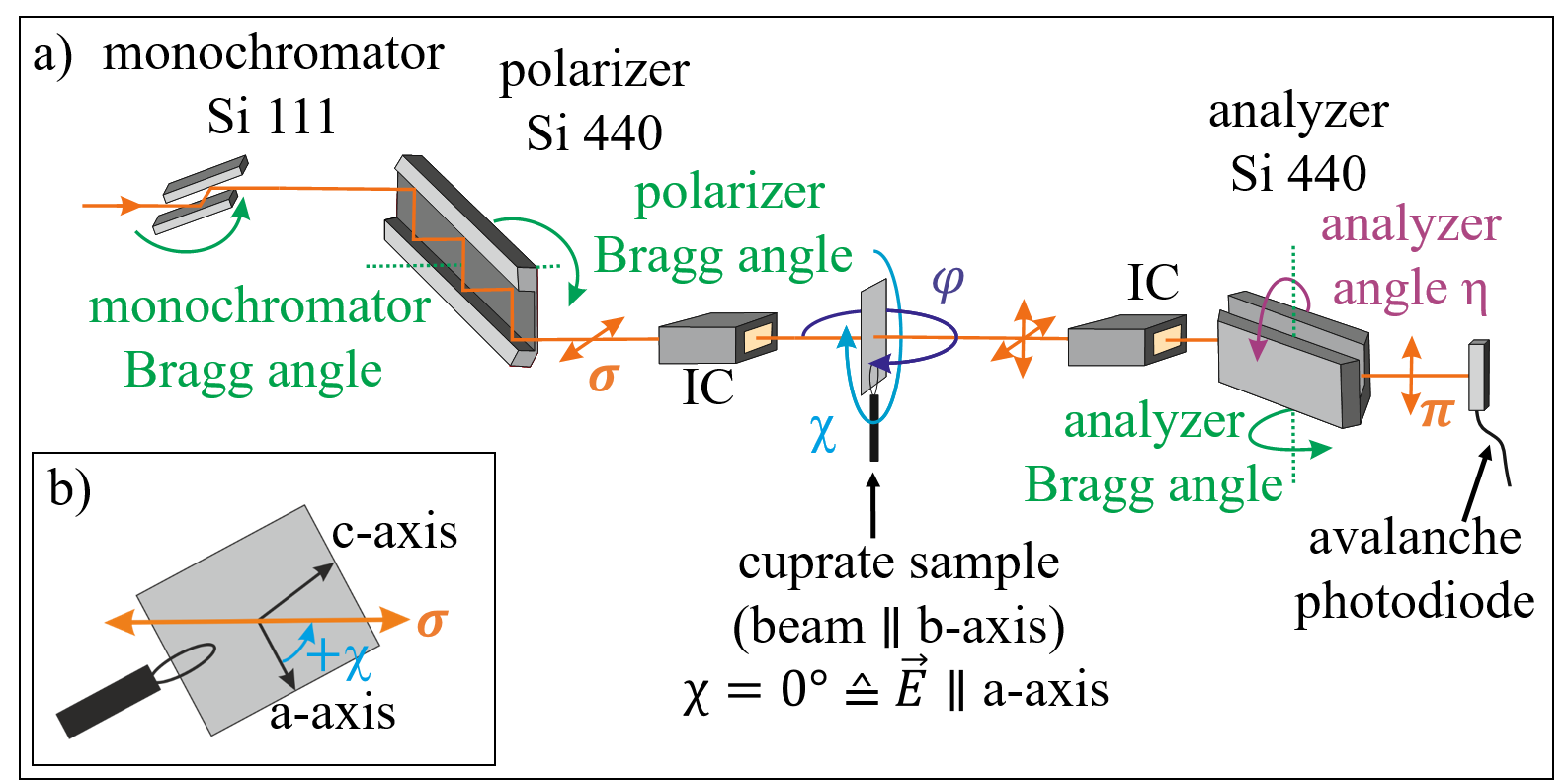}
		\caption{a) The sample was mounted on a Eulerian cradle with the $b$-axis parallel to the linearly polarized beam ($\sigma-\text{polarization}$ in the horizontal plane). An avalanche photodiode behind the X-ray analyzer in crossed position to the polarizer detected the $\sigma\rightarrow\pi$ scattered photons. Simultaneously ionization chambers (IC) measured the transmitted intensity through the sample. b) Enlarged view of the sample as seen from the direction of the incoming beam. }
		\label{Setup}
	\end{figure}	

\section{Experimental Procedures}	
\subsection{High-purity Spectropolarimetry}
	The experiments were performed at the synchrotron radiation source \text{PETRA III} (DESY, Hamburg) with the setup shown in Fig.\,\ref{Setup}a. A Si(111) heat-load monochromator selected a 1-eV wide energy band out of the incident radiation. The sample was located under ambient conditions between two monolithic Si(440) channel-cut crystals, acting as highly efficient linear polarizers due to multiple reflections under the near-$45^\circ$ Bragg angle for the X-ray energy of the Cu K-edge. In a $90^\circ$ crossed setting, these two crystals form a polarimeter consisting of polarizer and analyzer \cite{Hart_1991}, providing polarization extinction ratios of up to 10$^{-10}$ \,\,\cite{Marx_2013} or better \cite{Kai_2018}, which is way superior to what is possible in the regime of optical wavelengths. Any X-ray optical activity occuring between polarizer and analyzer converts a fraction of the highly pure $\sigma-\text{polarization}$ incident on the sample into the orthogonal $\pi-\text{polarization}$ that is then very efficiently transmitted by the analyzer and detected by an avalanche photodiode in single-photon counting mode. For precision angular adjustment relative to the incident beam, the sample was mounted on an Eulerian cradle, providing angular degrees of freedom along $\varphi$ and $\chi$.
	
	The energy resolution of the X-ray polarimeter is determined by the Darwin width of the Si(440) polarizer and analyzer rocking curves which translates here into an energy bandpass of \SI{62}{meV}, thus allowing for the detection of very sharp spectral features. For scanning the polarimeter over the energy range of the Cu K-absorption edge with a constantly high polarization purity, the Bragg angles of polarizer and analyzer crystals have to be varied simultaneously with the Bragg angle of the Si(111) monochromator.  
	To cover the energy range of the Cu-K-edge from \SI{8970}{eV} to \SI{9010}{eV}, the Bragg angle on the Si(440) plane has to be varied from $\theta_B=45.78^\circ$ to $\theta_B=46.06^\circ$. In this energy range we achieved a polarization purity better than $1.3\cdot10^{-8}$, which was measured at \SI{8970}{eV} and lies within the range between $1\cdot10^{-7}$ and $5\cdot10^{-9}$ that is predicted by the dynamical theory of X-ray diffraction.

\begin{figure*}[ht!]
	\centering\includegraphics[width=14cm]{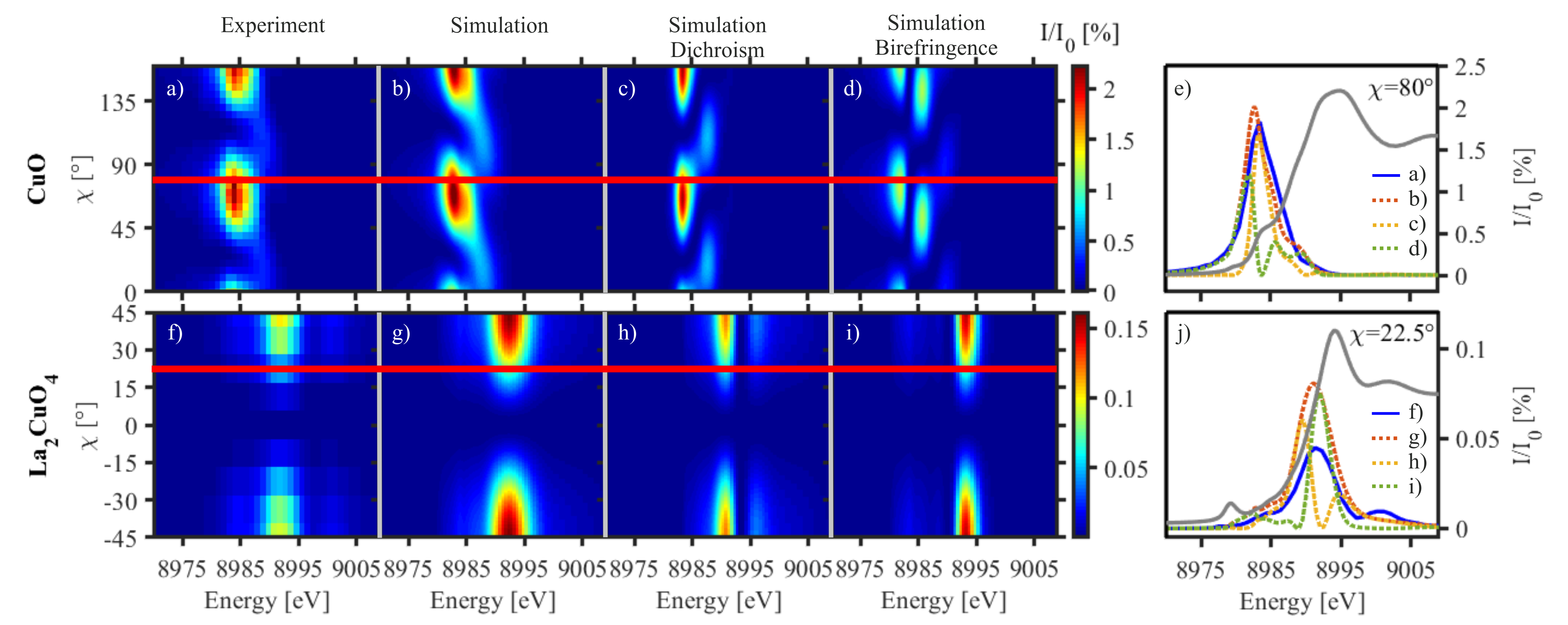}
		\caption{Intensity of the $\sigma \rightarrow \pi$ scattered photons of CuO (top) and La$_2$CuO$_4$ (bottom) normalized to the incident intensity $I_0$ on the sample for different angles $\chi$ between the crystal $a$-axis and the electric field vector. X-ray birefringence and dichroism were simulated by neglecting the anisotropic part of the real or imaginary part of the complex linear absorption coefficient $\mu$, respectively. The red line in a) to d) and f) to i) marks the $\chi$-position of the line out shown in e) and j), respectively. The corresponding XANES spectra are indicated in grey in e) and j).}
		\label{Messung}
	\end{figure*}

	To accurately determine the intensity of the $\sigma \rightarrow \pi$ scattered photons for a given polarimeter energy and angular setting ($\varphi, \chi$) of the sample (see Fig.\ref{Setup}),	the maximum of the rocking curve of the analyzer Bragg angle at each setting was taken. Fig.\,\ref{Messung} shows the measured spectra of the $\sigma \rightarrow \pi$ scattered photons for both crystals in the setting $\varphi = 0$ as a function of the angle $\chi$. By comparison with the conventional XANES spectra (see \textcolor{urlblue}{Supplement 1} and the grey lines in Figs.\,\ref{Messung}e,\,j), one observes that in case of CuO, the $\sigma \rightarrow \pi$ scattering is maximal at the position of the pre-edge peak at 8984 eV, while in case of La$_2$CuO$_4$ this maximum occurs at the inflection point of the absorption edge at 8994 eV. These energy dependencies can be related to the Cu orbital configuration in the respective compound as will be discussed later.
		
\subsection{Modelling the Complex Linear Absorption Coefficient}
	
		The dependence of X-ray dichroism and X-ray birefringence on the photon polarization and the sample orientation is ruled by the point group of the crystal \cite{Brouder_1990}. This is explained in detail in \textcolor{urlblue}{Supplement 1}. According to these considerations we chose the orientation $\varphi=0$ for both sample materials, which corresponds to the electric field vector lying in the $a$-$c$-plane. This allows to detect the full anisotropy of the electric dipole absorption cross-section $\sigma^{D}$.	Accordingly, the single crystals were shaped into $(010) - \text{orientated}$ slabs controlled by the Laue method with an accuracy of $\leq 0.3^\circ$. Subsequently they were gently polished to 33\,$\mu$m (CuO) and 23.5\,$\mu$m (La$_2$CuO$_4$) thin disks.  The thickness of the samples was determined by transmission measurements and comparison to Henke data  \cite{Henke_1993}.  Finally, the adjustment of the crystal axes within the $a$-$c$ plane was confirmed again by the Laue method.

	For a description of the optical activity of these samples, i.e., the $\sigma \rightarrow \pi$ scattering, we use the complex linear absorption coefficient \cite{Collins_2012, Joly_2012}

	\begin{align}
		\mu=\mu'+i\mu''=\begin{pmatrix}
		\mu_{\sigma\sigma} & \mu_{\sigma\pi}\\
		\mu_{\pi\sigma} & \mu_{\pi\pi}
		\end{pmatrix}
		\text{ , $\mu_{xx} \in \mathbb{C}$, }
	\end{align}

 The real part $\mu'$ is responsible for dichroism, whereas the imaginary part $\mu''$ is related to birefringence. 
	
In order to determine $\mu$ for the spectral region of the Cu K-edge of our samples, ab initio calculations were performed with the FDMNES code following the local density approximation \cite{Joly2001}. The relativistic full potential approach was used, including the spin-orbit interaction and the core-hole potential effect. Self-consistent electronic structures around the absorbing atom were calculated in a cluster with a radius of up to \SI{6}{\mathring{A}}. The code includes all calculation steps of the polarization changes in a material up to the final transmitted intensity after the analyzer.

For the calculation of polarization changes in the presence of a complex linear absorption coefficient, the Jones matrix formalism is used. The derivation is given in detail in \textcolor{urlblue}{Supplement 1}. Consequently, the $\pi-\text{polarized}$ X-ray intensity after the sample normalized to the impinging $\sigma-\text{polarized}$ X-ray intensity on the sample, $I_{\sigma\pi}$, is given by

	\begin{align}\label{scatter_formular}
		I_{\sigma\pi} = e^{-\frac{1}{2}\left(\mu_{\sigma\sigma}'+  \mu_{\pi\pi}'\right)l} \frac{|\sinh\left(\tau l\right)|^2}{8|\tau|^2} \times
		|\left(\mu_{\pi\pi}-\mu_{\sigma\sigma}\right)\sin 2\chi-2\mu_{\sigma\pi}\cos 2 \chi|^2,
	\end{align}

where $\mu_{\sigma\pi}=\mu_{\pi\sigma}$ for centrosymmetric crystals, and $\tau=\frac{1}{4}\sqrt{(\mu_{\pi\pi}-\mu_{\sigma\sigma})^{2} + 4\mu_{\sigma\pi}\mu_{\pi\sigma} }$ \cite{Lovesey_2001, Joly_2012}. $l$ is the thickness of the sample. 	

\section{Results}	

	Based on this theoretical description, we will now discuss the influence of the symmetry of the complex linear absorption coefficient $\mu$ on the measured $\sigma \rightarrow \pi$ scattered intensity $I_{\sigma\pi}$. According to equation (\ref{scatter_formular}), the following behavior is expected: If the non-diagonal tensor elements $\mu_{\pi\sigma}$ are zero,  $I_{\sigma\pi}$ is maximal at $\chi=\pm45^\circ$ for all energies, since $I_{\sigma\pi}$ is then proportional to $\sin^2 2\chi$. If the non-diagonal tensor elements $\mu_{\pi\sigma}$ are non-zero, the maxima of $I_{\sigma\pi}$ as a function of $\chi$ depend on the components of $\mu$ and thus are energy-dependent.
	
	Both symmetry cases of the complex linear absorption coefficient $\mu$  were experimentally verified by investigating the two samples La$_2$CuO$_4$ ($\mu_{\pi\sigma}=0$) and CuO ($\mu_{\pi\sigma}\neq0$), as shown in Fig.\,\ref{Messung}.  In agreement with the theory, all spectral features of La$_2$CuO$_4$ have the same angular dependence for which  $I_{\sigma\pi}$ is maximum at \mbox{$\chi=\pm45^\circ$}, whereas for CuO, there is an energy-dependent shift in $\chi$ for $I_{\sigma\pi}$ due to the nonzero $\mu_{\pi\sigma}$. Furthermore, $I_{\sigma\pi}$ has a $\chi-\text{periodicity}$ of $\pi/2$ for both crystals, which can easily be explained by their centrosymmetry. 
	The theoretical simulation is in very good qualitative agreement with the experimental data. This is further exemplified by line cuts at selected $\chi$ angles shown in Figs.\,\ref{Messung}e, j. While the agreement in case of CuO is excellent, the peak intensity for La$_2$CuO$_4$ is overestimated by the simulation.
	
	The theoretical description also allows to calculate the spectra of the $\sigma\rightarrow\pi$ transmission separately for X-ray birefringence and X-ray dichroism. This can be achieved by neglecting the anisotropic part of either the real part or the imaginary part of $\mu$, respectively, which is illustrated in Figs.\,\ref{Messung}c and \ref{Messung}d for CuO and Figs.\,\ref{Messung}h and \ref{Messung}i for La$_2$CuO$_4$. It turns out that the total $\sigma \rightarrow \pi$ scattered intensity $I_{\sigma\pi}$ is the linear superposition of the $\sigma \rightarrow \pi$ scattered intensities due to X-ray dichroism and birefringence. 

\begin{figure}
		\centering\includegraphics[width=0.55\textwidth]{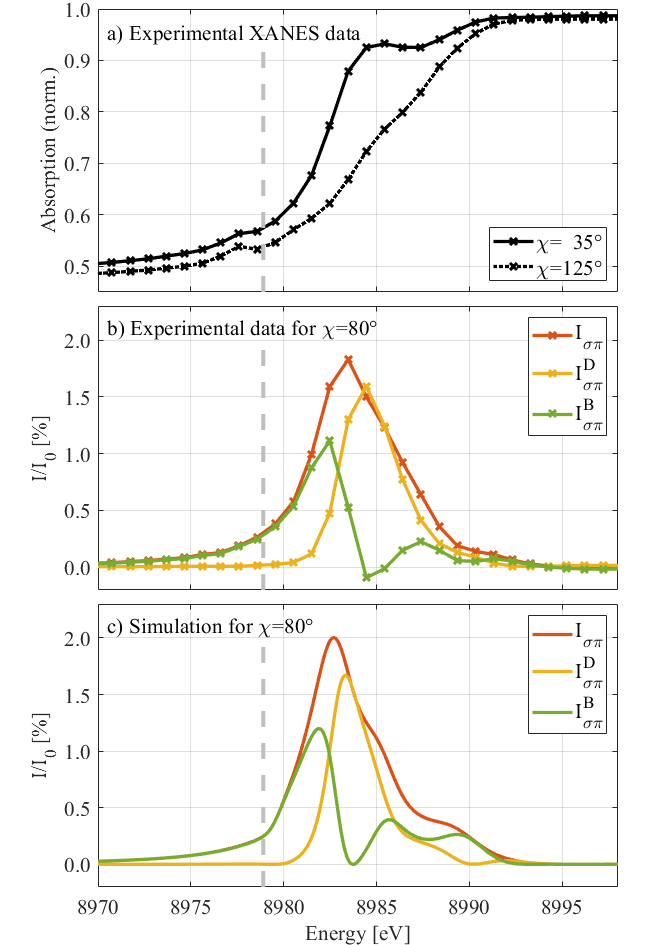}
		\caption{Experimentally disentangling dichroism and birefringence. a) The transmission ($=1-\text{absorption}$) through the sample at two orthogonal sample orientations $\chi\pm\frac{\pi}{4}$ is used to calculate the scattered intensity due to X-ray dichroism, $I_{\sigma\pi}^{D}(\chi)$, which can be subtracted from the total $I_{\sigma\pi}(\chi)$ to obtain the $\sigma \rightarrow \pi$ scattered intensity due to birefringence, $I_{\sigma\pi}^{B}(\chi)$, shown in b).c) Simulation with FDMNES. The dashed grey line marks the position of the Cu K-edge (\SI{8978.9}{eV}) according to Henke data \cite{Henke_1993}.}
		\label{Disentangle}
	\end{figure}

To verify this disentanglement of dichroism and birefringence experimentally, we have performed conventional X-ray linear dichroism measurements by taking XANES spectra for two orthogonal sample orientations. This is shown in Fig.\,\ref{Disentangle}a. For the calculation of the $\sigma \rightarrow \pi$ scattered intensity due to X-ray dichroism measured at the sample orientation $\chi$, $I_{\sigma\pi}^{D}(\chi)=(1/2)\sin^2\left[\pi/4 - \arctan\left(\sqrt{T^{+}/T^{-}}\right)\right]\cdot\left(T^{+}+T^{-}\right)$,  the transmission through the sample, $T^{\pm}=T(\chi\pm\frac{\pi}{4})$, measured at the sample orientations $\chi\pm\frac{\pi}{4}$ is needed. This formula can be derived very easily by vector superposition as explained in \textcolor{urlblue}{Supplement 1}. The $\sigma \rightarrow \pi$ scattered intensity due to X-ray birefringence can now the separated by a simple subtraction $I_{\sigma\pi}^{B}(\chi)=I_{\sigma\pi}(\chi)-I_{\sigma\pi}^{D}(\chi)$, which is plotted for $\chi=80^\circ$ in Fig.\,\ref{Disentangle}b, being in excellent agreement with the simulated data shown in Fig.\,\ref{Disentangle}c.

The polarization changes due to dichroism and birefringence can be attributed to specific projections of the density of states on the absorbing atoms. In case of investigating a (010) - oriented sample with an X-ray beam parallel to the crystal $b-\text{axis}$, the density of states $\delta(p_x)$, $\delta(p_z)$, $\delta(d_{xy})$ and $\delta(d_{yz})$ of the orbitals $p_x$, $p_z$, $d_{xy}$ and $d_{yz}$ are involved, where $z$ is choosen along the $c-\text{axis}$ of the crystal. $I_{\sigma\pi}$ is related to the difference between the density of states $\delta(p_x)-\delta(p_z)$ and $\delta(d_{xy})-\delta(d_{yz})$. It turned out that $I_{\sigma\pi}^{D}$ is maximal, when $\delta(p_x)-\delta(p_z)$ is maximal, while $I_{\sigma\pi}^{B}$ is maximal for energies where $\delta(p_x)\approx\delta(p_z)$. This is illustrated in \textcolor{urlblue}{Supplement 1}, Fig.\,3.

The X-ray absorption cross section of CuO and La$_2$CuO$_4$ at the Cu K-edge is highly dominated by dipole transitions (E1E1). Consequently, the simulation of the  $\sigma \rightarrow \pi$ transmission spectra for different multipole contributions with FDMNES showed that they are mainly due to dipole (E1E1) transitions with a three orders of magnitude weaker quadrupole (E2E2) contribution caused by the 1s$\rightarrow$3d pre-peak of the absorption spectrum. This is presented in \textcolor{urlblue}{Supplement 1}, Fig.\,4. Mixed dipole-quadrupole (E1E2) and dipole electric-magnetic (E1M1) transitions did not play a role at the Cu K-edge for CuO and La$_2$CuO$_4$.
		  
\begin{figure}
		\centering\includegraphics[width=0.7\textwidth]{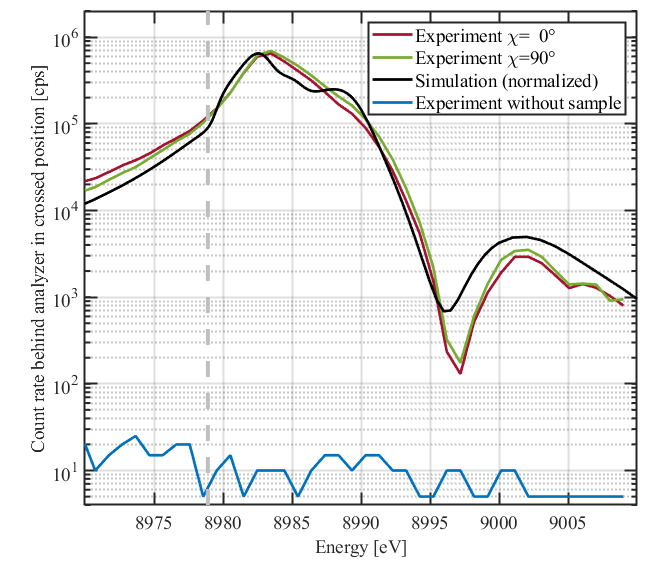}
		\caption{Intensity behind the analyzer in crossed position for $\chi=0^\circ$ between the $a$-axis of the CuO crystal and the electric field vector of the beam. To demonstrate the reproducibility of the measurement, the data for $\chi=90^\circ$ are also shown, which is in good agreement for the simulation over many orders of magnitude. The dashed grey line marks the position of the Cu K-edge (\SI{8978.9}{eV}) according to Henke data \cite{Henke_1993}.}
		\label{Dynamic}
	\end{figure}

The analysis of $I_{\sigma\pi}$ has considerable advantages over conventional X-ray absorption measurements such as X-ray natural linear dichroism (XNLD): It is essentially background free and allows to monitor X-ray optical activity over a dynamic range of several orders of magnitude. Fig.\,\ref{Dynamic} shows the $\sigma \rightarrow \pi$ transmission of the CuO sample for $\chi = 0^{\circ}$ and $\chi = 90^{\circ}$. This measurement not only proves the $90^{\circ}$ periodicity of the anisotropy in this sample, but also highlights the excellent agreement with the theoretical simulation over almost four orders of magnitude. An energy scan that was performed without a sample in the beam illustrates the very low background level of the polarimeter. Thus, a dynamical range of six orders of magnitude and essentially background-free measurements enable unprecedented sensitivity to detect optical activity in the energy range of the Cu K-edge from \SI{8970}{eV} to \SI{9010}{eV}. 
This renders our technique particularly sensitive to higher-order transitions such as weak 1s$\rightarrow$3d quadrupolar excitations that are located in the pre-edge region. These transitions often display a pronounced linear dichroism and birefringence due to the symmetry of the orbitals in the excited state \cite{Hahn1982,Glatzel2009,Cabaret2010}. Since they are excited from a spherically symmetric 1s ground state, they provide an attractive spectroscopic signature to probe the 3d orbitals of the valence shell. This could be particularly attractive for the study of cuprate superconductors which owe their properties to the electronic structure of their CuO$_2$ planes. Revealing the anisotropy and occupancy of the corresponding Cu orbitals and their hybridizations, could thus provide clues about the origin of superconductivity in the cuprates. Moreover, adressing the K-edges in the regime of hard x-rays allows one to access samples in absorbing environments like high-pressure cells or buried layers in thin-film systems and induces less radiation damage compared to soft x-rays. In conventional x-ray absorption spectroscopies, pre-edge features in transition metal compounds are often overshadowed by the strong 1s$\rightarrow$4p transition, which is the reason why they are only rarely studied. The suppression of any isotropic scattering by the crossed polarizers that leads to the exceptional signal-to-noise ratio of this method will alleviate the latter limitation, thus providing a new approach to obtain clear views on charge and orbital anisotropies in the valence shell.
	
\section{Discussion and Conclusion}

This work reports the first comprehensive experimental and theoretical investigation of X-ray birefringence and dichroism at the Cu K-edge for two different crystal systems. By measuring the X-ray dichroism conventionally, the real part of the complex linear absorption coefficient can be determined. The imaginary part, which corresponds to X-ray birefringence, can be determined by subtracting the measured dichroism spectra from the $\sigma \rightarrow \pi$ scattered photon spectra. This is especially interesting for the determination of optical constants of materials that cannot be treated via ab initio calculations. Examples are materials that contain impurities like those that have been modified by ion implantation or doping, and those very strongly correlated systems, for which no suitable theoretical approach is currently capable of simulating their properties.

High purity polarimetry can adress a large part of the K- absorption edges (elements between $Z=17$ and $Z=43$) and L- absorption edges (elements between $Z=37$ and $Z=92$) by a suitable crystal reflection with an Bragg angle near $45^\circ$. Corresponding polarimeters made of silicon, germanium or quartz can reach in most cases theoretically a polarization purity of $<10^{-10}$ (\textcolor{urlblue}{Supplement 1}, Tables\,1-4). This opens a wide field of application potential for the investigation of electronic anisotropies via high purity polarimetry. One example for a highly interesting research field could be the investigation of iron based superconductors at the Fe K-edge with a Si(133) polarimeter.
	
This method has several advantages over conventional methods for the detection of dichroism, such as XNLD, and can answer questions of fundamental importance. It can avoid the problem of integrating via a finitely measured absorption cross-section using the Kramers-Kronig relation in order to obtain the real part of the refractive index, which is the conventional approach. High polarization sensitivity is particularly suitable for observing small anisotropies as early indicators of phase transitions during or even long before reaching critical parameters. This new approach is especially interesting for investigating very weak anisotropies of quadrupolar or octopolar transitions in the pre-edge region like they were recently detected in Gd$_3$Ga$_5$O$_{12}$ at the Gd L$_1$-absorption edge \cite{Amelie_2019}. In contrast to XNLD, this method does not require spectra of orthogonal orientations to be subtracted from each other. Instead, the measurement of anisotropies with a high angular resolution is directly and quickly accessible. This enables single-shot measurements for time-resolved investigations of certain spectral features in a pump-probe setting, for example. Furthermore, in analogy to an optical polarization microscope, in combination with micro-focused x-ray beams it is also possible to map and image X-ray polarization anisotropies with very high spatial resolution. 

Disentangling dichroism and birefringence with high sensitivity will also play a pivotal role in future experiments on probing QED in extreme electromagnetic fields with polarized X-rays. The experiments proposed so far \cite{Heinzl2006, karbstein2015vacuum, karbstein2018vacuum, karbstein2016probing, karbstein2020enhancing, schlenvoigt2016detecting} aim at the detection of the birefringence of the vacuum. 
A realistic scenario of such an experiment, originally proposed in \cite{Heinzl2006}, has been quantitatively worked out in \cite{karbstein2018vacuum}: A pulse of linearly polarized hard x-rays (12914 eV) from an x-ray laser traverses a 1$\mu$m wide focal waist of a laser beam ($\lambda$ = 800\,nm) with 30\,J pulse energy and 30\,fs pulse duration. In this interaction, assuming $N = 10^{12}$ photons in the x-ray pulse, QED calculations predict a small single-digit number of x-ray photons to flip their polarization from $\sigma$ to $\pi$. These photons can be detected by a high-purity polarimeter of the kind described here with an extinction ratio in the range of 10$^{-12}$ that is reachable today \cite{Kai_2018}.
A dichroic contribution to the signal would imply a correction to QED. It would, therefore, be a direct hint for the existence of particles beyond the standard model such as millicharged or axion-like particles \cite{Gies06,Maiani86}.


\section*{Funding}
This work was funded by the Deutsche Forschungsgemeinschaft (DFG) under Grant No. 416700351 within the Research Unit FOR2783/1, by the Bundesministerium f\"ur Bildung und Forschung (BMBF) under Grant No. 05K16SJ2 and by the European Social Fund (ESF) and the Free State of Thuringia in the framework of Forschergruppe 2017FGR0074. 

\section*{Acknowledgement}
We would like to thank Heike Marschner and Ortrud Wehrhan for sharing their expertise in crystallography, and Claudia R\"odl, Martin von Zimmermann, Paul Schenk and Felix Karbstein for helpful and enlightening discussions. Parts of this research were carried out at \text{PETRA III}, and we would like to thank Ilya Sergeev, Hlynur Gretarsson, Rene Steinbr\"ugge, Frank-Uwe Dill and Conrad Hagemeister for assistance in using beamline P01. 

\section*{Disclosures}
The authors declare no conflicts of interest.
\medskip

See \textcolor{urlblue}{Supplement 1} for supporting content.

\bibliography{Literature_Spectropol}
	
\end{document}


\title{Supplemental material}

\author{
	Annika T. Schmitt,\authormark{1,2,3,*} 
	Yves Joly,\authormark{4}
	Kai S. Schulze,\authormark{1,2,3}
	Berit Marx-Glowna,\authormark{1,2,3}
	Ingo Uschmann, \authormark{1,2,3}
	Benjamin Grabiger, \authormark{1,2,3}
	Hendrik Bernhardt, \authormark{1,2,3}
	Robert Loetzsch, \authormark{1,2,3}
	Am\'elie Juhin, \authormark{5}
	J\'er\^ome Debray, \authormark{4}
	Hans-Christian Wille, \authormark{6}
	Hasan Yava\c{s}, \authormark{6,7}
	Gerhard G. Paulus, \authormark{1,2,3}
	and Ralf R\"ohlsberger \authormark{6,1,2,3,**}
}

\address{
	\authormark{1}Institut f\"ur Optik und Quantenelektronik, Friedrich-Schiller-Universit\"at Jena, Max-Wien-Platz 1, 07743 Jena, Germany\\
	\authormark{2}Helmholtz-Institut Jena, Fr\"{o}belstieg 3, 07743 Jena, Germany\\
	\authormark{3}Helmholtz Centre for Heavy Ion Research (GSI), Planckstr. 1, 64291 Darmstadt, Germany\\
	\authormark{4}Institut N\'eel, 25 Rue des Martyrs, BP 166, 38042 Grenoble cedex 09, France\\
	\authormark{5}Institut de Min\'eralogie, de Physique des Mat\'eriaux et de Cosmochimie (IMPMC), Sorbonne Universit\'e, UMR 7590, 4 place Jussieu, 75252 Paris Cedex 05, France\\
	\authormark{6}Deutsches Elektronen-Synchrotron DESY, Notkestr.\,85, 22607 Hamburg, Germany\\
	\authormark{7}SLAC National Accelerator Laboratory, 2575 Sand Hill Road, MS103, Menlo Park, CA 94025
	
	\authormark{*}annika.schmitt@uni-jena.de
	\authormark{**}ralf.roehlsberger@desy.de
}

\begin{abstract}
This document provides supplemental information to "Disentangling X-ray dichroism and birefringence via high-purity polarimetry". We give details on the theoretical description of the $\sigma\rightarrow\pi$ scattered intensity,  the relation between the complex linear absorption coefficient and the atomic scattering factor, and on the conventional near-edge x-ray absorption spectra of CuO and La$_2$CuO$_4$. Moreover, we explain the relation between the observed dichroism and birefringence with the densities of states related to the electronic orbitals of the compounds.
\end{abstract}

\section{Angular dependence of the X-ray absorption cross-section $\sigma$}

The samples in this experiment belong to two different point groups: 

\begin{itemize} 
\item monoclinic CuO with point group $C_{2h}(2/m)$ and space group $C2/c$ \\
($a$ = 4.6837\,\AA, $b$ = 3.4226\,\AA, $c$ = 5.1288\,\AA, and $\beta=99.54^\circ$ \cite{Asbrink_1969})  and 
\item orthorhombic La$_{2}$CuO$_{4}$ with point group $D_{2h}(mmm)$ and space group $Bmab$ \\
($a$ = 5.352\,\AA, $b$ = 5.400\,\AA, $c$ = 13.157\,\AA, and $\beta=90^\circ$ \cite{Tuilier_1992}). 
\end{itemize}

Both materials show trichroism, which means that all eigenvalues of the absorption cross-section tensor $\sigma^D$ are different. The electronic transitions of CuO and La$_2$CuO$_4$ with the main absorption edge at approximately \SI{8994}{eV} are mainly of electric dipole origin. The electric dipole absorption cross-section is given by
	
	\begin{align}
	\sigma^{D}(\hat{\epsilon})=\sigma^{D}(0,0)-\sqrt{\frac{8\pi}{5}}\sum\limits_{m=-2}^{2}Y_2^{m*}(\hat{\epsilon})\ \sigma^{D}(2,m)
	\end{align}
	
which depends (a) on the polarization vector $\hat{\epsilon}$ and the sample orientation relative to the photon wavevector through the spherical harmonics $Y^m_l$ and (b) on the point group of the crystal and the photon energy through the tensor $\sigma^D$ of the dipole absorption cross-section,
with the complex tensor components $\sigma^{D}(l,m)$ \cite{Brouder_1990}, where $m$ and $l$ are the angular momentum quantum numbers.
$\sigma^{D}(0,0)$ is the isotropic part of the electric dipole absorption cross-section which is independent of the polarization. 
The polarization vector $\hat\epsilon=\left[\sin(\chi-\beta ) \cos\varphi,\ \sin(\chi-\beta ) \sin\varphi,\  \cos(\chi-\beta ) \right]$ is chosen in accordance with the experimental setup where $\chi$ represents the angle between the $a$-axis of the crystal and the electric field vector of the synchrotron beam as shown in Fig.\,1b. For $\chi=0^\circ$, the electric field vector is parallel to the $a$-axis. $\beta$ is the base angle of the crystal lattice which is $\neq 90^{\circ}$ for the oblique lattice of monoclinic CuO.\\
	
	In the case of CuO with point group $C_{2h}(2/m)$, where $\sigma^{D}(2,1)=\sigma^{D}(2,-1)=0$, the electric dipole absorption cross-section is given by
	
	\begin{align}
	\begin{split}
	\sigma^{D}(\hat{\epsilon}) = &\sigma^{D}(0,0)-\sqrt{3}\sin^2(\chi-\beta )[\cos2\varphi \operatorname{\mathbb{R}e}\{\sigma^{D}(2,2)\}+ \sin2\varphi \operatorname{\mathbb{I}m}\{\sigma^{D}(2,2)\}] \\ 
	                             &-\frac{1}{\sqrt{2}}\left[3\cos^2(\chi-\beta )-1\right]\sigma^{D}(2,0).
	\end{split}
	\label{dipole_cross_section_CuO}
	\end{align}
	
	\noindent For La$_{2}$CuO$_{4}$ with point group $D_{2h}(mmm)$, where $\sigma^{D}(2,1)=\sigma^{D}(2,-1)=0$ and $\sigma^{D}(2,2)=\sigma^{D}(2,-2)$, $\sigma^{D}(\hat{\epsilon})$ is simplified  to
	
		\begin{align}
	\begin{split}
	\sigma^{D}(\hat{\epsilon}) = \
	\sigma^{D}(0,0)-\sqrt{3}\sin^2(\chi-\beta )\cos2\varphi \operatorname{\mathbb{R}e}\{\sigma^{D}(2,2)\} 
	-\frac{1}{\sqrt{2}}\left[3\cos^2(\chi-\beta )-1\right]\sigma^{D}(2,0).
	\end{split}
	\end{align}

\section{Theoretical description of \boldmath$\sigma - \pi$ scattered intensity}
	
\noindent We use the $(2\times2)$ Jones Matrix defined in a $(\sigma, \pi)$ basis, $\sigma$ horizontal, $\pi$ vertical, and the wave vector $k$ perpendicular to this basis. 

\subsection{Transmittance matrix}

\noindent We use the transmittance matrix under the $\sigma - \pi$ basis where $\sigma$ and $\pi$ are 2 directions perpendicular to the wave vector. We note $\mu$ the complex linear absorption coefficient. Following Lovesey and Collins \cite{Lovesey_2001}, after a distance $l$, the transmittance matrix is given by:

\begin{align}
T(l)=e^{-\frac{1}{4}(\mu_{\sigma\sigma}+\mu_{\pi\pi})l}\begin{pmatrix}
\cosh(\tau l) + \frac{\mu_{\pi\pi}-\mu_{\sigma\sigma}}{4\tau}\sinh\left(\tau l\right)& -\frac{\mu_{\sigma\pi}}{2\tau}\sinh\left(\tau l\right) \\
-\frac{\mu_{\pi\sigma}}{2\tau}\sinh\left(\tau l\right) & \cosh\left(\tau l\right) + \frac{\mu_{\sigma\sigma}-\mu_{\pi\pi}}{4\tau}\sinh\left(\tau l\right)
\end{pmatrix}
\end{align}

\begin{align*}
\text{with } \tau=\frac{1}{4}\sqrt{(\mu_{\pi\pi}-\mu_{\sigma\sigma})^{2} + 4\mu_{\sigma\pi}\mu_{\pi\sigma} }=\frac{1}{4}(\mu_{\pi\pi}-\mu_{\sigma\sigma})\sqrt{1+\frac{4\mu_{\sigma\pi}\mu_{\pi\sigma}}{(\mu_{\pi\pi}-\mu_{\sigma\sigma})^{2}}}
\end{align*}

\subsection{Polarization matrix}

\noindent In the following we also use the Jones matrix for the polarization:

\begin{align}
P=\frac{1}{2}\begin{pmatrix}
1+S_{1} & S_{2}-i S_{3}\\
S_{2}+i S_{3} & 1-S_{1}
\end{pmatrix}.
\end{align}

\noindent When rotating the sample by an angle $\chi$ the linear polarization gets an angle $-\chi$ versus its $\sigma$ direction. Thus one has:

\begin{align}
S_{1}=\cos2\chi, S_{2}=-\sin 2\chi, S_{3}=0
\end{align}

\begin{align}
P=\frac{1}{2}\begin{pmatrix}
1+\cos2\chi & -\sin2\chi\\
-\sin2\chi & 1-\cos2\chi
\end{pmatrix} =
\begin{pmatrix}
\cos^2\chi & -\sin\chi\cos\chi\\
-\sin\chi\cos\chi & \sin^2\chi
\end{pmatrix}.
\end{align}

\subsection{Analyzer matrix}

\noindent The analyzer matrix (or polarizer after the sample) is given versus its rotation angle, $\eta$, and its Bragg angle, $\theta$, by:

\begin{align}
A=\begin{pmatrix}
\cos\eta & -\sin\eta\\
\cos 2\theta\sin\eta & \cos 2\theta\cos\eta
\end{pmatrix}.
\end{align}

\noindent For a perfect analyzer crystal $\cos2\theta=0$. For the polarization flip by $90^\circ$, one then obtains:

\begin{align}
A=\begin{pmatrix}
\cos\left(-\frac{\pi}{2}+\chi\right) & -\sin\left(-\frac{\pi}{2}+\chi\right)\\
0 & 0
\end{pmatrix}=
\begin{pmatrix}
\sin\chi & \cos\chi\\
0 & 0
\end{pmatrix}.
\end{align}

\subsection{Transmitted intensity}

\noindent After the path through the sample, the intensity is given by

\begin{align}
I=Tr\left(ATPT^\dagger A^\dagger\right)
=|\sin\chi \cos\chi (T_{\sigma\sigma}-T_{\pi\pi})+\cos^2\chi T_{\pi\sigma}-\sin^2\chi T_{\sigma\pi} |^2
\end{align}

\noindent Non-diagonal non-magnetic case $T_{\sigma\pi}=T_{\pi\sigma}$: 

\begin{align}
I=|\sin\chi \cos\chi (T_{\sigma\sigma}-T_{\pi\pi})+\cos2\chi T_{\pi\sigma}|^2
\end{align}

\begin{align}
\begin{pmatrix}
T_{\sigma\sigma}& T_{\sigma\pi}\\
T_{\pi\sigma}&T_{\pi\pi}
\end{pmatrix}=e^{-\frac{1}{4}\left(\mu_{\sigma\sigma}+\mu_{\pi\pi}\right)l}\begin{pmatrix}
\cosh(\tau l)+\frac{\mu_{\pi\pi}-\mu_{\sigma\sigma}}{4\tau}\sinh(\tau l) & -\frac{\mu_{\sigma\pi}}{2\tau}\sinh(\tau l)\\
-\frac{\mu_{\pi\sigma}}{2\tau}\sinh(\tau l) & \cosh(\tau l)+ \frac{\mu_{\sigma\sigma}-\mu_{\pi\pi}}{4\tau}\sinh(\tau l)
\end{pmatrix},
\end{align}

\begin{align}
\begin{split}
 I=& e^{-\frac{1}{2}\left(\mu_{\sigma\sigma}'+\mu_{\pi\pi}'\right)l}|\sin\chi \cos\chi\left(\frac{\mu_{\pi\pi}-\mu_{\sigma\sigma}}{2\tau}\sinh(\tau l)\right)-\cos2\chi\frac{\mu_{\sigma\pi}}{2\tau}\sinh(\tau l)|^2\\=& e^{-\frac{1}{2}\left(\mu_{\sigma\sigma}'+\mu_{\pi\pi}'\right)l} \frac{|\sinh(\tau l)|^2}{8|\tau|^2}|(\mu_{\pi\pi}-\mu_{\sigma\sigma})\sin2\chi-2\mu_{\sigma\pi}\cos2\chi|^2
 \end{split}
 \end{align}
\noindent with
\begin{align}
	\tau=\frac{1}{4}\sqrt{(\mu_{\pi\pi}-\mu_{\sigma\sigma})^2+4\mu_{\sigma\pi}\mu_{\pi\sigma}}
\end{align} 
 
\newpage

\section{Derivation of the relation between complex linear absorption coefficient, complex refractive index and complex atomic scattering factor}

\noindent A plane wave described by

\begin{align}
	E(z,t)=E_{0}e^{i\left(kz-\omega t\right)}
\end{align}

\noindent passes through a medium with complex refractive index $n=1-\delta+i\beta=\frac{c\cdot k}{\omega}$:

\begin{align}
	E(z,t)&=E_{0}e^{i[\frac{\omega}{c}(1-\delta+i\beta)z-\omega t]}\\
	&=E_{0}e^{i\omega\left(\frac{z}{c}-t\right)}\cdot e^{-\frac{2\pi}{\lambda}\left(i\delta+\beta\right)z}.
	\label{Brechungsindex}
\end{align}

\noindent By introducing the complex linear absorption coefficient $\mu=\mu'+i\mu''$, following S.P. Collins et al. \cite{Collins_2012}, the expression can be written as

\begin{align}
	E(z,t)=\tilde{E_{0}}\cdot e^{-\mu\frac{z}{2}}
	\label{Absorptionskoeffizient}
\end{align}

\noindent with the vacuum propagation $\tilde{E_0}=E_{0}e^{i\omega\left(\frac{z}{c}-t\right)}$. This leads to the Lambert-Beer law for the intensity $I=|E|^2$:

\begin{align}
	I=I_{0} e^{-\mu'z}.
\end{align}

\noindent A comparison of eq. \ref{Brechungsindex} and eq. \ref{Absorptionskoeffizient} shows that the complex linear absorption coefficient $\mu=\mu'+i\mu''$ is connected to the refractive index $n=1-\delta+i\beta$ via
\begin{align}
	&\mu'=\frac{4\pi}{\lambda}\beta \text{ and}\\
	&\mu''=\frac{4\pi}{\lambda}\delta.
\end{align}

\noindent Since the complex refractive index $n$ can also be expressed by using the complex atomic scattering factor $f=f'-if''$,

\begin{align}
	n=1-\frac{n_{a}r_{e}\lambda^2}{2\pi}(f'-if'')=1-\delta+i\beta,
\end{align}

\noindent where $n_a$ is the atomic density and $r_{e}$ is the classical electron radius, there is also a relation between the complex linear absorption coefficient and the complex atomic scattering factor:

\begin{align}
	\mu'=2n_{a}r_{e}\lambda f''\\
	\mu''=2n_{a}r_{e}\lambda f'.
\end{align}

\section{X-ray absorption near edge structure of CuO and La$_2$CuO$_4$}

\begin{figure}[ht]
	\centering\includegraphics[width=0.95\textwidth]{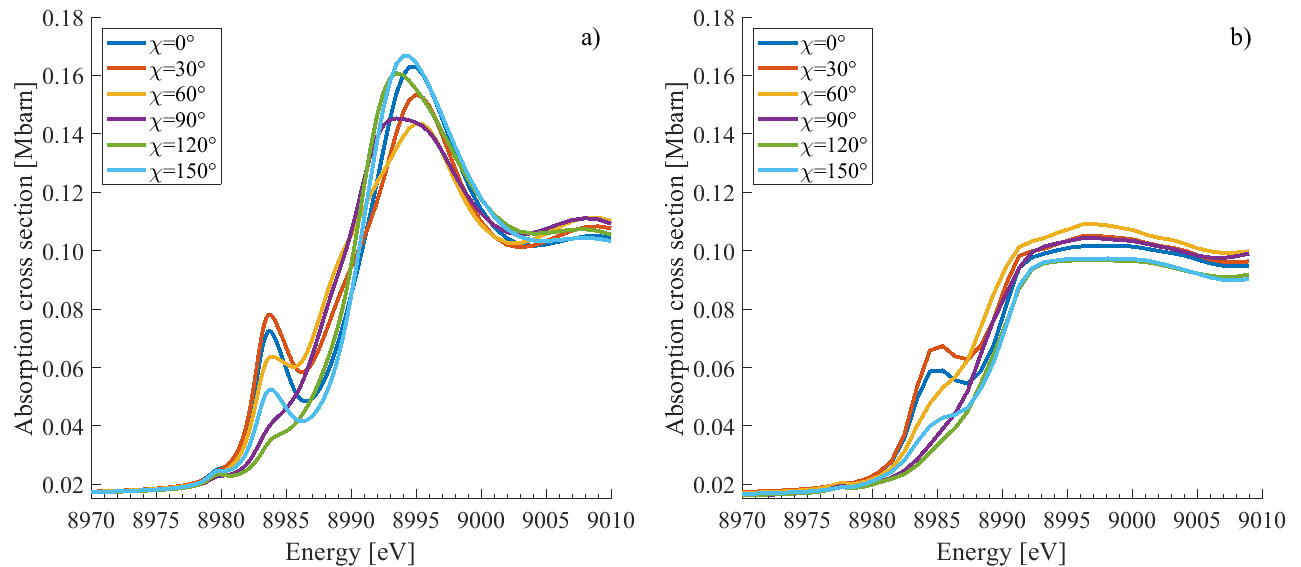}
	\centering\includegraphics[width=0.95\textwidth]{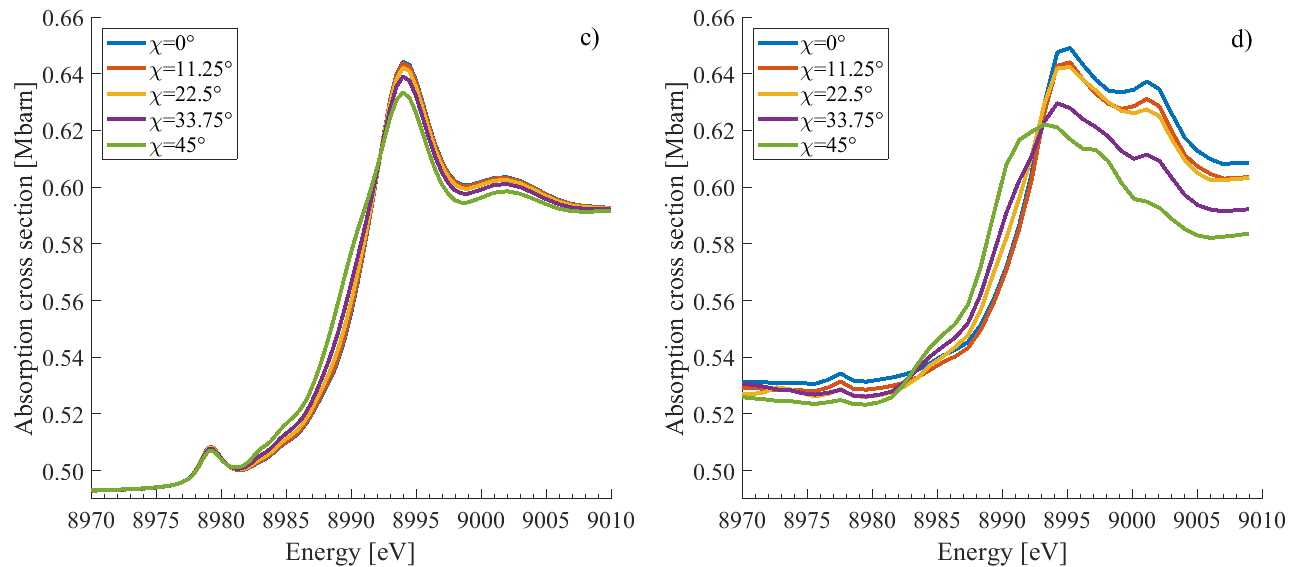}
	\caption{X-ray absorption near edge structure (XANES) simulated for CuO (a) and for La$_2$CuO$_4$ (c) for different sample orientations $\chi$. Experimental data for CuO and La$_2$CuO$_4$ are shown in (b) and (d), respectively.}
	\label{XANES}
\end{figure}

The absorption cross-section was simulated for CuO (Fig.\,\ref{XANES}a) and La$_2$CuO$_4$ (Fig.\,\ref{XANES}c) with the ab-initio quantum code FDMNES in units of Mbarn ($\SI{1}{Mbarn}=10^{-18}\text{cm}^2$) for different angles $\chi$ between the electric field vector and the $a-\text{axis}$ of the crystal \cite{Joly2001}. It was also determined experimentally by measuring the transmission through the sample with ionization chambers for CuO (Fig.\,\ref{XANES}b) and La$_2$CuO$_4$ (Fig.\,\ref{XANES}d). The main absorption edge (white line) is at approximately \SI{8994}{eV}. 

\section{Calculation of  the $\sigma \rightarrow \pi$ scattered intensity due to X-ray dichroism}

The $\sigma \rightarrow \pi$ scattered intensity due to X-ray dichroism can be calculated using experimental XANES data, which are measured by ionization chambers before and behind the sample. Two orthogonal transmission measurements at the sample orientations $\chi\pm\frac{\pi}{4}$ are needed to calculate the $\sigma \rightarrow \pi$ scattered intensity due to X-ray dichroism at the sample orientation $\chi$, which is normalized to the incident intensity $I_{in}$. The incoming electric field vector $E_{in}$ is rotated by an angle $\Delta\chi$ due to dichroism in the sample, i.e. the different transmission of the electric field, $t$, at the sample orientations $\chi\pm\frac{\pi}{4}$:  

\begin{align}
	\Delta\chi=\frac{\pi}{4}-\arctan\left(\frac{t(\chi+\frac{\pi}{4})}{t(\chi-\frac{\pi}{4})}\right)
\end{align}

\begin{figure}[ht]
	\centering\includegraphics[width=0.35\textwidth]{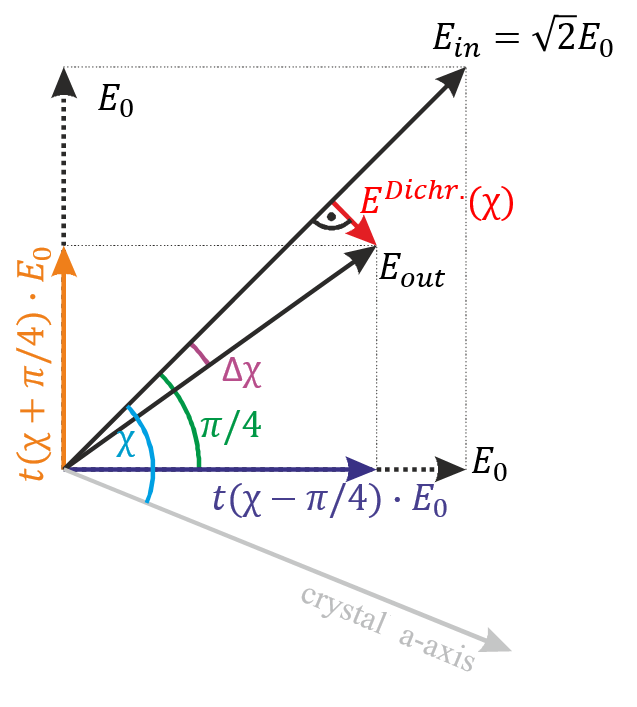}
	\caption{The incident electric field vector $E_{in}$ is rotated by an angle $\Delta\chi$ due to the different transmission along the two orthogonal axis at the sample orientations $\chi\pm\frac{\pi}{4}$. The outgoing electric field vector $E_{out}$ is no longer purely $\sigma-\text{polarized}$, but has now a $\pi-\text{component}$, $E^{Dichr.}\left(\chi\right)$, which can pass the X-ray analyzer in crossed position to the polarizer. The $\sigma \rightarrow \pi$ scattered photons due to X-ray dichroism at the sample orientation $\chi$, $E^{Dichr.}\left(\chi\right)$ , can  be easily calculated when the transmitted intensity at the sample orientations $\chi\pm\frac{\pi}{4}$, $t^2(\chi\pm\frac{\pi}{4})\cdot E_0^2$, is measured simultaneously with ionization chambers before and behind the sample, while the sample is rotated.}
	\label{Dichroism}
\end{figure}

The incident electric field vector $E_{in}$ is $\sigma-\text{polarized}$. The outgoing electric field vector 

\begin{align}
	E_{out}=\left[t^2(\chi+\frac{\pi}{4})\cdot E_0^2+t^2(\chi-\frac{\pi}{4})\cdot E_0^2\right]^{1/2}
\end{align}

 is rotated and therefore has also a $\pi-\text{polarization component}$, which can pass the analyzer in crossed position to the polarizer and is measured behind the analyzer as $\sigma \rightarrow \pi$ scattered photons $E^{Dichr.}(\chi)$. This can be calculated by

\begin{align}\label{E_Dichroism}
E^{Dichr.}(\chi)=\sin\Delta\chi\cdot E_{out}
\end{align}

The $\sigma \rightarrow \pi$ scattered intensity due to dichroism, $E_{Dichr.}^2(\chi)$, is normalized to the incident intensity $I_{in}=\left(\sqrt{2}E_0\right)^2$:

\begin{align}
I_{\sigma\pi}^{D}(\chi)=\frac{E_{Dichr.}^2}{\left(\sqrt{2}E_0\right)^2}(\chi)=\frac{1}{2}\sin^2\left[\frac{\pi}{4}-\arctan\left(\sqrt{\frac{T^{+}}{T^{-}}}\right)\right]\cdot \left(T^{+}+T^{-}\right)
\end{align}

with the transmitted intensity $T^{\pm}=T(\chi\pm\frac{\pi}{4})=t^2(\chi\pm\frac{\pi}{4})$.

\section{Density of states of CuO and La$_2$CuO$_4$}

The density of states were simulated with FDMNES. $z$ was choosen along $c-\text{axis}$ of the crystal. For a $(010)-\text{oriented}$ sample and a beam along the crystal $b-\text{axis}$, the involved orbitals are $p_x$, $p_z$, $d_{xy}$ and $d_{yz}$. A comparison between the simulation of the $\sigma \rightarrow \pi$ scattered photons and the density of states shows, that the intensity of $\sigma \rightarrow \pi$ scattered photons due to dichroism is maximal for energies with a maximal difference between the density of states of $p_x$ and $p_z$. In the other hand, the intensity of $\sigma \rightarrow \pi$ scattered photons due to birefringence is maximal for energies where the density of states of $p_x$ and $p_z$ are equal.

\begin{figure}[h!]
	\includegraphics[width=0.92\textwidth]{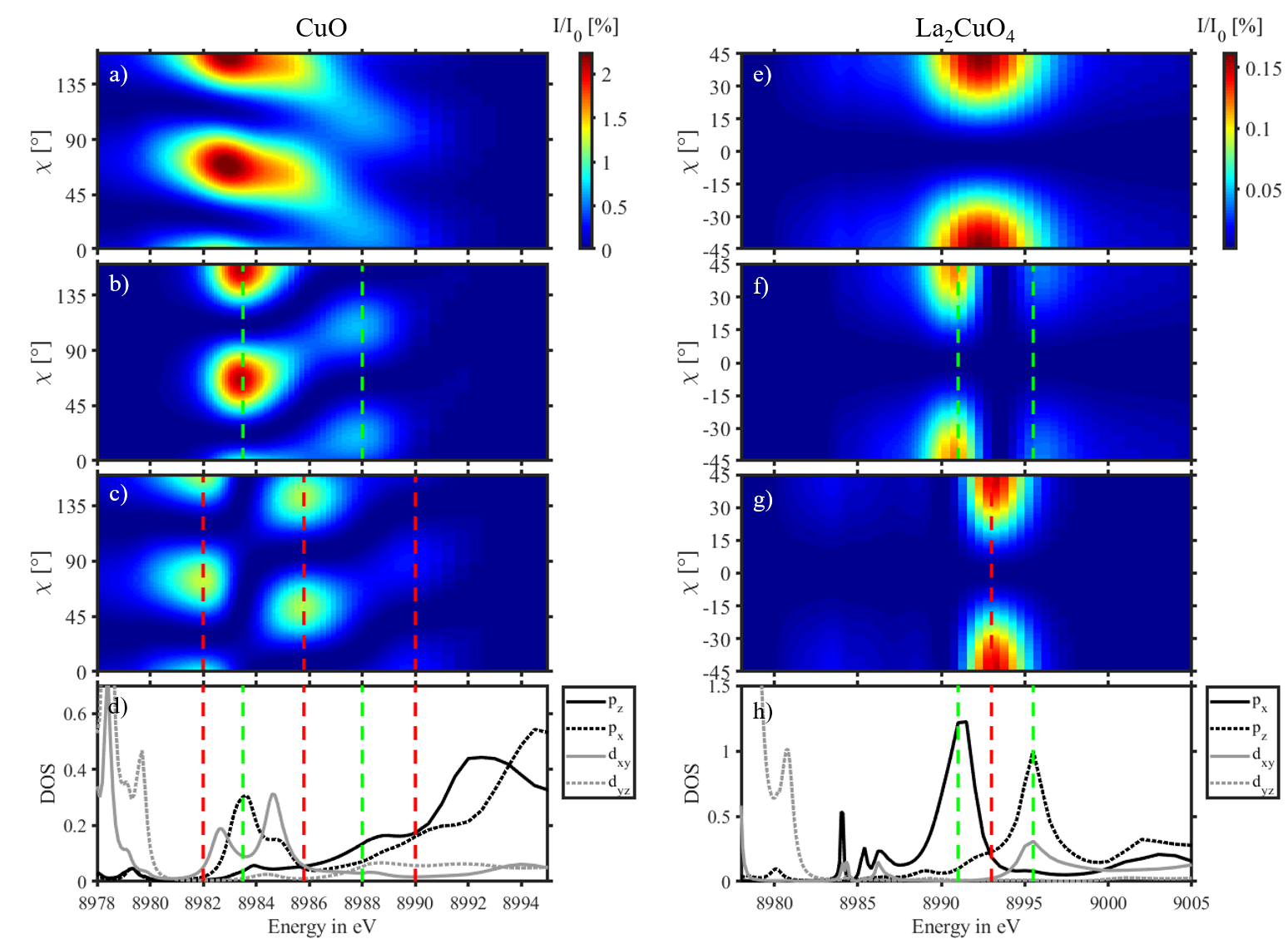}
	\caption{ a) Simulated intensity of the $\sigma \rightarrow \pi$ scattered photons of CuO and b) for La$_2$CuO$_4$ for different angles $\chi$ between the crystal $a$-axis and the electric field vector. The intensity of the $\sigma \rightarrow \pi$ scattered photons X-ray due to dichroism b) and f), and due to birefringence c) and g) can be attributed to differences between the density of states $\delta(p_x)-\delta(p_z)$ and $\delta(d_{xy})-\delta(d_{yz})$, respectively. The density of states of the orbitals $p_x$, $p_z$, $d_{xy}$ and $d_{yz}$ are shown for CuO in d) and for La$_2$CuO$_4$ in h). Maximal difference between the density of $p-\text{states}$ leads to dichroism (green lines), while birefringence is maximal at energies, where the density of states of both $p-\text{Orbitals}$ are equal (red lines).}
	\label{DOS}
\end{figure}
\newpage

\section{Contribution of multipole transitions to spectra}

The intensity of the $\sigma \rightarrow \pi$ scattered photons and the X-ray absorption near edge structure spectra were simulated with FDMNES for electric dipole (E1E1), quadrupole (E2E2), dipole-qudrupole (E1E2) and dipole electric-magnetic (E1M1) transitions separately. The electric dipole term highly dominates the $\sigma \rightarrow \pi$ scattered photons and the absorption cross section of CuO and La$_2$CuO$_4$. The electric quadrupole contribution (E2E2) to the $\sigma \rightarrow \pi$ scattered photons is three orders of magnitute weaker than the electric dipole for both elements. Mixed electric dipole-quadrupole transitions and dipole electric-magnetic transitions do not contribute to the spectra.

\begin{figure}
	\centering
	\includegraphics[width=1\textwidth]{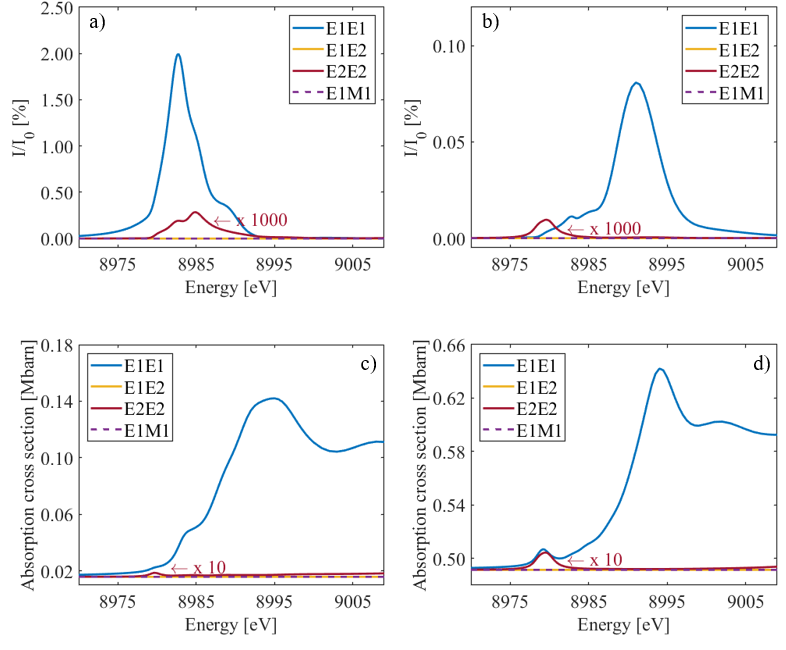}
	\caption{a) Simulated intensity of the $\sigma \rightarrow \pi$ scattered photons of CuO at $\chi=80^\circ$ and b) for La$_2$CuO$_4$ at $\chi=22.5^\circ$ c) X-ray absorption near edge structure (XANES) simulated for CuO at $\chi=80^\circ$ and for d) La$_2$CuO$_4$ at $\chi=22.5^\circ$. Electric dipole (E1E1), quadrupole (E2E2), dipole-qudrupole (E1E2) and dipole electric-magnetic (E1M1) transitions were calculated separately.}
	\label{Fig10}
\end{figure}
\newpage

\section{Absorption edges addressable via high-purity polarimetry}

A large part of the X-ray absorption edges can be addressed using high-precision X-ray polarimetry. An overview of possible X-ray polarimeters for the K absorption edges is tabulated in Table\,\ref{Table1}, and for the L absorption edges in  Table\,\ref{Table2}, \,\ref{Table3} and \ref{Table4}. The corresponding polarization purities were determined by calculating the rocking curve for $\pi- $ and $\sigma-\text{polarization}$ components at the corresponding energy using dynamical theory of X-ray diffraction and considering the number of reflections in the channel-cut. The polarization purity is then obtained from the ratio of the $\pi$ and $\sigma$ components. For silicon and germanium, which can be realized as conventional channel-cut crystals, 6 reflections per channel-cut were assumed for the calculation of the polarization purity. For quartz and diamond, which can be used as quasi channel-cuts \cite{Bernhardt2020}, 4 reflections were used. The polarimeter option, which allows the highest possible polarization purity for the specific energy, is tabulated. The calculation does not take into account detour excitations and must be clarified for each individual case. Only crystal reflections with a FWHM of at least 0.1 arcsec were considered.

 \renewcommand{\arraystretch}{0.9}

\begin{table}[h!]
		\centering
	\caption{Possible crystal reflections for an X-ray polarimeter for the energies of the K absorption edge with a polarization purity that can be expected according to dynamic theory with 6 reflections per channel-cut for Si and Ge and 4 reflections per channel-cut for quartz and diamond. }
	\label{Table1}
\begin{tabular}[h]{lccr}
\hline
\\[-1em]
\multicolumn{4}{c}{K absorption edge}	\\
\hline
\\[-1em]	
Element &	Energy [eV] &	Polarimeter	& Polarization puriy\\
\hline
\\[-1em]
17  Cl &	2822.4 &	Si (1 1 1) &	<1.0E-12\\
19  K	& 3608.4 &	quartz (-2 1 0) &	<1.0E-12\\
20  Ca	& 4038.5 &quartz (0 2 0)& 3.4E-12\\
21  Sc	& 4492	& Ge (0 2 2)	   & <1.0E-12\\
22  Ti	& 4966	& quartz (1 1 2)  &	1.7E-06\\
23  V	& 5465	& Si (1 1 3)    &	1.1E-10\\
24  Cr	& 5989	& quartz (-1 2 -3)&	5.0E-12\\
25  Mn	& 6539	& Si (0 0 4)&	9.2E-11\\
26  Fe	& 7112	& Si (1 3 3)&	1.5E-12\\
27  Co	& 7709	& Ge (2 2 4)&	2.3E-10\\
28  Ni	& 8333	& Si (1 1 5)&	<1.0E-12\\
29  Cu	& 8979	& Si (4 4 0)&	5.8E-08\\
30  Zn	& 9659	& Si (3 1 5)&	5.7E-11\\
31  Ga	& 10367	& Ge (3 3 5)&	2.0E-10\\
32  Ge	& 11103	& Si (4 4 4)&	2.2E-11\\
33  As	& 11867	& Ge (2 4 6)&	<1.0E-12\\
34  Se	& 12658	& Ge (0 0 8)&	<1.0E-12\\
35  Br	& 13474	& Ge (2 2 8)&	<1.0E-12\\
36  Kr	& 14326	& Ge (0 4 8)&	<1.0E-12\\
37  Rb	& 15200	& Ge (4 4 8)&	<1.0E-12\\
38  Sr	& 16105	& Ge (0 2 10)&	<1.0E-12\\
39  Y	& 17038	& Ge(2 4 10)&	<1.0E-12\\
40  Zr	& 17998	& Ge (0 8 8)&	<1.0E-12\\
41  Nb	& 18986	& Ge (0 0 12)&	<1.0E-12\\
42  Mo	& 20000	& Ge (0 4 12)&	<1.0E-12\\
43  Tc	& 21044	& Ge  (4 4 12)&	<1.0E-12\\
\hline
\end{tabular}
\end{table}

\begin{table}[h!]
	\centering
	\caption{Possible crystal reflections for an X-ray polarimeter for the energies of the L$_1$ absorption edge with a polarization purity that can be expected according to dynamic theory with 6 reflections per channel-cut for Si and Ge and 4 reflections per channel-cut for quartz and diamond. }
		\label{Table2}
	\begin{tabular}[h]{lccr}
		\hline
		\\[-1em]
		\multicolumn{4}{c}{L$_1$ absorption edge}	\\
		\hline
		\\[-1em]
		Element &	Energy [eV] &	Polarimeter	& Polarization puriy\\
		\hline
		\\[-1em]
	37  Rb&	2065&	quartz (-1 1 0)&<1.0E-12\\
	41  Nb&	2698&	Ge (1 1 1)&	<1.0E-12\\
	42  Mo&	2866&	Si (1 1 1)&	<1.0E-12\\
	46  Pd&	3604&	quartz (1 1 0)&	<1.0E-12\\
	47  Ag&	3806&	quartz (1 -1 2)&<1.0E-12\\
	48  Cd&	4018&	quartz (0 2 0)&2.3E-11\\
	49  In&	4238&	diamond (111)&<1.0E-12\\
	50  Sn&	4465&	Ge (0 2 2)&<1.0E-12\\
	51  Sb&	4698&	quartz (1 1 2)&7.6E-09\\
	52  Te&	4939&	quartz (1 1 2)&3.7E-07\\
	53  I&	5188&	Ge (1 1 3)&	<1.0E-12\\
	54  Xe&	5453&	Si (1 1 3)&	2.9E-11\\
	55  Cs&	5714&	quartz (3 -2 -1)&1.2E-12\\
	56  Ba&	5989&	quartz (-1 2-3)&5.0E-12\\
	57  La&	6266&	Ge (0 0 4)&	1.3E-12\\
	58  Ce&	6549&	Si (0 0 4)&	3.4E-10\\
	59  Pr&	6835&	Ge (1 3 3)&	<1.0E-12\\
	60  Nd&	7126&	Si (3 3 1)&	1.1E-11\\
	61  Pm&	7428&	quartz (-1 4 0)&<1.0E-12\\
	62  Sm&	7737&	Ge (2 2 4)&	2.5E-09\\
	63  Eu&	8052&	Ge (1 1 5)&	<1.0E-12\\
	64  Gd&	8376&	Si (1 1 5)&	<1.0E-12\\
	65  Tb&	8708&	Ge (0 0 4)&<1.0E-12\\
	66  Dy&	9046&	Si (0 4 4)&1.1E-10\\
	67  Ho&	9394&	Ge (1 3 5)&	2.9E-09\\
	68  Er&	9751&	Ge (0 2 6)&	<1.0E-12\\
	69  Tm&	10116&	Si ( 0 2 6)&1.9E-10\\
	70  Yb&	10486&	Ge (4 4 4)	&1.6E-08\\
	71  Lu&	10870&	Ge (4 4 4)	&6.1E-12\\
		72  Hf&	11271&	Ge (2 4 6)&	<1.0E-12\\
		73  Ta&	11682&	Ge (2 4 6)&	<1.0E-12\\
		74  W&	12100&	Si (2 4 6)&	<1.0E-12\\
		75  Re&	12527&	Ge (0 0 8)&	<1.0E-12\\
		76  Os&	12968&	Si (0 0 8)&	<1.0E-12\\
		77  Ir&	13419&	Ge (2 2 8)&	<1.0E-12\\
		78  Pt&	13880&	Ge (0 4 8)	&<1.0E-12\\
		79 Au&	14353&	Ge (0 4 8)&	<1.0E-12\\
		80  Hg&	14839&	Ge (4 6 6)&	<1.0E-12\\
		81  Tl&	15347&	Ge (4 4 8)&	<1.0E-12\\
		82  Pb&	15861&	Si (4 4 8)&	<1.0E-12\\
		83  Bi&	16388&	Ge (5 1 9)&	<1.0E-12\\
		84  Po&	16939&	Ge (2 4 10)&	<1.0E-12\\
		85  At&	17493&	Ge (2 4 10)&	<1.0E-12\\
		86  Rn&	18049&	Ge (0 8 8)	&<1.0E-12\\
		87  Fr&	18639&	Ge (6 8 8)	&<1.0E-12\\
		88  Ra&	19237&	Ge (0 0 12)	&<1.0E-12\\
		89  Ac&	19840&	Ge (0 4 12)	&<1.0E-12\\
		90  Th&	20472&	Ge (2 8 10)	&<1.0E-12\\
		91  Pa&	21105&	Ge (4 4 12)	&<1.0E-12\\
		\hline	
	\end{tabular}
\end{table}

\begin{table}[h!]
	\centering
	\caption{Possible crystal reflections for an X-ray polarimeter for the energies of the L$_2$ absorption edge with a polarization purity that can be expected according to dynamic theory with 6 reflections per channel-cut for Si and Ge and 4 reflections per channel-cut for quartz and diamond. }
		\label{Table3}
	\begin{tabular}[h]{lccr}
		\hline
		\\[-1em]
		\multicolumn{4}{c}{L$_2$ absorption edge}	\\	
		\hline
		\\[-1em]
		Element &	Energy [eV] &	Polarimeter	& Polarization puriy\\
		\hline
		\\[-1em]
		38  Sr&	2007&	quartz (0 1 0)&	6.4E-11\\
		42  Mo&	2625&	quartz (1 0 1)&	<1.0E-12\\
		43  Tc&	2793&	Si (1 1 1)&	<1.0E-12\\
		47  Ag&	3524&	quartz (1 1 0)&	<1.0E-12\\
		48  Cd&	3727&	quartz (1 -1 2)&1.7E-10\\
		49  In&	3938&	quartz (1 -1 2)	& 1.5E-11\\
		50  Sn&	4156&	quartz (0 2 0)&	<1.0E-12\\
		51  Sb&	4380&	Ge (0 2 2)&	<1.0E-12\\
		52  Te&	4612&	Si (0 2 2)&	<1.0E-12\\
		53  I&	4852&	quartz (1 1 2)&	8.5E-12\\
		54  Xe&	5107&	Ge (1 1 3)&	<1.0E-12\\
		55  Cs&	5359&	Si (1 1 3)&	<1.0E-12\\
		56  Ba&	5624&	quartz (3 -2 -1)&	8.5E-10\\
		57  La&	5891&	quartz (-2 3 1)&	7.0E-06\\
		58  Ce&	6164&	Ge (0 0 4)&	<1.0E-12\\
		59  Pr&	6440&	Si (0 0 4)&	<1.0E-12\\
		60  Nd&	6722&	Ge (1 1 3)&	<1.0E-12\\
		61  Pm&	7013&	Si (1 3 3)&	<1.0E-12\\
		62  Sm&	7312&	quartz (-2 -2 0)&	5.1E-06\\
		63  Eu&	7617&	Ge (2 2 4)&	<1.0E-12\\
		64  Gd&	7930&	Si (2 4 2)&	<1.0E-12\\
		65  Tb&	8252&	Si (1 1 5)&	9.8E-10\\
		66  Dy&	8581&	Ge (0 4 4)&	1.2E-08\\
		67  Ho&	8918&	Ge (0 4 4)& 1.3E-09\\
		68  Er&	9264&	Ge (1 3 5)&	<1.0E-12\\
		69  Tm&	9617&	Si (1 3 5)&	<1.0E-12\\
		70  Yb&	9978&	Ge (0 2 6)&	1.7E-09\\
		71  Lu&	10349&	Ge (3 3 5)&	7.1E-11\\
		72  Hf&	10739&	Ge (4 4 4)&	<1.0E-12\\
		73  Ta&	11136&	Si (4 4 4)&	<1.0E-12\\
		74  W&	11544&	Ge (2 4 6)&	<1.0E-12\\
		75  Re&	11959&	Ge (2 4 6)&	<1.0E-12\\
		76  Os&	12385&	Ge (0 0 8)&	<1.0E-12\\
		77  Ir&	12824&	Ge (0 0 8)&	<1.0E-12\\
		78  Pt&	13273&	Ge (2 2 8)&	<1.0E-12\\
		79 Au&	13734&	Si (2 2 8)&	<1.0E-12\\
		80  Hg&	14209&	Ge (0 4 8)&	<1.0E-12\\
		81  Tl&	14698&	Ge (4 6 6)&	<1.0E-12\\
		82  Pb&	15200&	Si (4 6 6)&	<1.0E-12\\
		83  Bi&	15711&	Ge (4 4 8)&	<1.0E-12\\
		84  Po&	16244&	Ge (0 2 10)&<1.0E-12\\
		85  At&	16785&	Ge (2 4 10)&<1.0E-12\\
		86  Rn&	17337&	Ge (2 4 10)&<1.0E-12\\
		87  Fr&	17907&	Ge (0 8 8)&	<1.0E-12\\
		88  Ra&	18484&	Ge (6 6 8)&	<1.0E-12\\
		89  Ac&	19083&	Ge (0 0 12)&<1.0E-12\\
		90  Th&	19693&	Ge (2 2 12)&<1.0E-12\\
		91  Pa&	20314&	Ge (0 4 12)&	<1.0E-12\\
		92  U&	20948&	Ge (4 4 12)&	<1.0E-12\\
		\hline	
	\end{tabular}
\end{table}

\begin{table}[h!]
	\centering
	\caption{Possible crystal reflections for an X-ray polarimeter for the energies of the L$_3$ absorption edge with a polarization purity that can be expected according to dynamic theory with 6 reflections per channel-cut for Si and Ge and 4 reflections per channel-cut for quartz and diamond. }
		\label{Table4}
	\begin{tabular}[h]{lccr}
		\hline
		\\[-1em]
		\multicolumn{4}{c}{L$_3$ absorption edge}	\\	
		\hline
		\\[-1em]	
		Element &	Energy [eV] &	Polarimeter	& Polarization puriy\\
		\hline
		\\[-1em]
		39  Y&	2080&	quartz (-1 1 0)&<1.0E-12\\
		43  Tc&	2677&	Ge (1 1 1)&<1.0E-12\\
		44  Ru&	2838&	Si (1 1 1)&	<1.0E-12\\
		48  Cd&	3538&	quartz (1 1 0)&	<1.0E-12\\
		49  In&	3730&	quartz (1 -1 2)&1.3E-10\\
		50  Sn&	3929&	quartz (1 -1 2)&7.2E-12\\
		51  Sb&	4132&	quartz (0 2 0)&	<1.0E-12\\
		52  Te&	4341&	Ge (0 2 2)&	<1.0E-12\\
		53  I&	4557&	Si (0 2 2)&	<1.0E-12\\
		54  Xe&	4786&	quartz (1 1 2)&	1.4E-11\\
		55  Cs&	5012&	Ge (1 1 3)&	<1.0E-12\\
		56  Ba&	5247&	Ge (1 1 3)&	<1.0E-12\\
		57  La&	5483&	quartz (-3 1 0)&<1.0E-12\\
		58  Ce&	5723&	quartz (3 -2 -1)&1.1E-11\\
		59  Pr&	5964&	quartz (-1 2 -3)&1.8E-10\\
		60  Nd&	6208&	Ge (0 0 4)&	<1.0E-12\\
		61  Pm&	6459&	Si (0 0 4)&	<1.0E-12\\
		62  Sm&	6716&	Ge (1 3 3)&	<1.0E-12\\
		63  Eu&	6977&	Si (1 1 3)&	<1.0E-12\\
		64  Gd&	7243&	Si (1 1 3)&	1.1E-07\\
		65  Tb&	7514&	Ge (2 2 4)&	2.0E-12\\
		66  Dy&	7790&	Si (2 4 2)&	5.2E-09\\
		67  Ho&	8071&	Ge (5 1 1)&	<1.0E-12\\
		68  Er&	8358&	Si (1 5 1)&	<1.0E-12\\
		69  Tm&	8648&	Ge (0 4 4)&	7.7E-11\\
		70  Yb&	8944&	Ge (0 4 4)&7.1E-09\\
		71  Lu&	9244&	Ge (1 3 5)&	<1.0E-12\\
		72  Hf&	9561&	Si (3 1 5)&	<1.0E-12\\
		73  Ta&	9881&	Ge (0 2 6)&	<1.0E-12\\
		74  W&	10207&	Si (0 2 6)&	<1.0E-12\\
		75  Re&	10535&	Ge (4 4 4)&	1.4E-09\\
		76  Os&	10871&	Ge (4 4 4)&	6.7E-12\\
		77  Ir&	11215&	Si (4 4 4)&	<1.0E-12\\
		78  Pt&	11564&	Ge (2 4 6)&	<1.0E-12\\
		79 Au&	11919&	Ge (2 4 6)&	<1.0E-12\\
		80  Hg&	12284&	Ge (0 0 8)&	<1.0E-12\\
		81  Tl&	12658&	Ge (0 0 8)&	<1.0E-12\\
		82  Pb&	13035&	Ge (2 2 8)&<1.0E-12\\
		83  Bi&	13419&	Ge (2 2 8)&<1.0E-12\\
		84  Po&	13814&	Ge (0 4 8)&	<1.0E-12\\
		85  At&	14214&	Ge (0 4 8)&	<1.0E-12\\
		86  Rn&	14619&	Ge (4 6 6)&	<1.0E-12\\
		87  Fr&	15031&	Ge (4 6 6)&	<1.0E-12\\
		88  Ra&	15444&	Ge (4 4 8)&	<1.0E-12\\
		89  Ac&	15871&	Si (4 8 4)&<1.0E-12\\
		90  Th&	16300&	Ge (0 2 10)&<1.0E-12\\
		91  Pa&	16733&	Ge (2 4 10)&<1.0E-12\\
		92  U&	17166&	Ge (2 4 10)&<1.0E-12\\
		\hline	
	\end{tabular}
\end{table}

\FloatBarrier
\newpage

\bibliography{Literature_Spectropol}